\documentstyle[12pt]{article}
\begin{document}

\def\lsim{\ \matrix{<\cr\noalign{\vskip-7pt}\sim\cr} \ }
\def\gsim{\ \matrix{>\cr\noalign{\vskip-7pt}\sim\cr} \ }
 
\begin{titlepage}

\hfill{PURD-TH-93-10}

\hfill{}

\hfill{July 1993}

\vskip 1cm

\centerline{\Large \bf Mini-charged tau neutrinos?}

\vskip 0.75cm

\centerline{\bf R. Foot$^{(a)}$ and H. Lew$^{(b)}$}

\vskip 0.75cm

\noindent
\centerline{{\it (a) Physics Department, McGill university, 
3600 University Street,}} 
\centerline{{\it Montreal, Quebec, H3A 2T8, Canada. }}
\centerline{{\it Email: foot@hep.physics.mcgill.ca}}

\vskip 2 mm
\noindent
\centerline{{\it (b) Physics Department, Purdue University,}}
\centerline{{\it West Lafayette, IN 47907-1396, U.S.A.}}
\centerline{{\it Email: lew\%purdd.hepnet@LBL.Gov}}
\vskip 1cm

\centerline{\large\bf Abstract}
\vskip 1cm
\noindent

Theoretically, the electric charge of the tau neutrino may be non-zero.
The experimental bound on the electric charge of the tau neutrino 
is many orders of magnitude weaker than that for any other 
known neutrino. If the tau neutrino does have a small electric charge, 
and its mass is greater than 1 MeV, then it can annihilate sufficiently 
in the early Universe by electromagnetic interactions to avoid conflict 
with the standard cosmology model. A novel feature of this scenario 
is that there can be effectively less than three neutrino species 
present during nucleosynthesis.

\end{titlepage}

The standard model contains 12 fermions -- 6 quarks and 6 leptons.
Theoretically, it is known that the electric charges of these fermions
are heavily constrained by the classical structure\footnote{The electric
charges are also constrained by anomaly cancellation requirements.
However, it might be possible  that the anomalies cancel in a trivial
manner (e.g. the existence of mirror fermions) leaving no constraints.
In this paper we are considering the option where anomaly constraints
play no role in electric charge quantization.} of the standard model 
\cite{cq,rf}. In fact there are just four classically undetermined 
electric charges in the standard model.
These four undetermined parameters can be taken to be
the charges of the 3 neutrinos and the charge of the down
quark ( we denote these four electric charges as
$Q(\nu_e)$, $Q(\nu_{\mu})$, $Q(\nu_{\tau})$ and $Q(d)$ ). 
All of the other fermion electric charges can be uniquely determined 
from the classical structure of the standard model in 
terms of these four classically undetermined parameters.
Experimentally, it is known that\footnote{We have chosen units such
that $Q(e) - Q(\nu_e) = 1.$}
\begin{eqnarray}
Q(d)& = &-{1 \over 3} \pm \delta_d, 
\quad \delta_d < 10^{-21} \nonumber \\
Q(\nu_e)& = & 0 \pm \delta_{\nu_e}, 
\quad \delta_{\nu_e} <  10^{-21} \nonumber \\
Q(\nu_{\mu})& = & 0 \pm \delta_{\nu_{\mu}}, 
\quad \delta_{\nu_{\mu}} <  10^{-9} \nonumber \\
Q(\nu_{\tau})& = & 0 \pm \delta_{\nu_{\tau}}, 
\quad \delta_{\nu_{\tau}} < 3\times 10^{-4},
\end{eqnarray}
where the experimental bounds on the $\delta$ parameters come from
experiments on the neutron charge \cite{bk}, the neutrality of 
matter \cite{mm}, $\nu_{\mu}e$ scattering \cite{bv} and the
SLAC beam dump experiment \cite{dcb}.

In addition to carrying a little electric charge, neutrinos
may also be massive. The experimental bounds on the masses of the 
neutrinos are given by \cite{pdb}:
\begin{eqnarray}
m(\nu_e) & < & 10 \hbox{ eV} \nonumber \\ 
m(\nu_{\mu})& < & 250 \hbox{ keV} \nonumber \\
m(\nu_{\tau})& < & 35 \hbox{ MeV} 
\end{eqnarray}
In addition to these direct experimental constraints, there are also other
constraints which can be obtained by demanding that the particle
physics be consistent with the standard cosmology model.
Although these ``bounds'' are not as dependable as the experimental
bounds due to the nature of cosmology (e.g. it involves untested
assumptions about the nature of the Universe a long time ago),
they are nevertheless still of interest since the standard 
cosmology model is simple and quite successful \cite{lv}. 
There are two important bounds from cosmology:
(1) The predicted energy density of massive neutrinos 
can a priori violate the observed bound.
(2) The number of light particles and their interactions 
can be constrained from nucleosynthesis.
If the neutrinos are stable and do not annihilate sufficiently
in the early Universe, then consistency with the standard cosmology
model requires that they must have mass less than about 100 eV, which
for the $\mu$ and $\tau$ neutrinos, is many orders of magnitude
less than the experimental bounds. If the electric charges of the 
neutrinos are zero, then they can only annihilate via weak interactions.
It has been shown some time ago \cite{lswein} ( see also \cite{others} ),
that they only annihilate sufficiently ( by weak interactions ) 
if their mass is greater than about 2 GeV. Since we know from experiments
that the known neutrinos all have mass less than 2 GeV, 
they will not annihilate sufficiently by weak interactions.
However, if they have electric charge, then they can annihilate 
via the electromagnetic interactions. We will now consider the case of 
a charged tau neutrino ( since the experimental bound on its charge is by 
far the weakest out of the three known neutrinos ). 

If the tau neutrino has a mass\footnote{ Since we are considering charged 
tau neutrinos, they must be Dirac fermions rather than Majorana ones
if they are massive.} greater than about 1 MeV,
then tau and anti-tau neutrinos can annhilate into electron and positron 
pairs via a s-channel photon.
The cross section for this process is quadratic in the tau neutrino electric
charge. For tau neutrino masses less than 1 MeV, the annihilation
cross section is considerably smaller since the
annihilation into electron positron pairs is kinematically forbidden.
The leading order contribution to the cross section is the
process $\nu_{\tau} \bar \nu_{\tau} \rightarrow \gamma \gamma$ 
and the cross section is proportional to the fourth power of the
tau neutrino charge. For this reason we will consider the region of
parameter space $m_{\nu_{\tau}} > 2 m_e \sim 1$ MeV.

To calculate the relic number density of the heavy tau neutrinos we
use the rate equation,
\begin{equation}
{dn \over dt} = -3 {{\dot R}\over R}n 
- \langle \sigma |v|\rangle n^2
+ \langle \sigma |v| \rangle n_0^2,
\label{rate}  
\end{equation}
where $n$ is the actual number density of heavy neutrinos at time
t, $n_0$ is the equilibrium number density, $R$ is the cosmic scale factor, 
$\langle \sigma |v|\rangle$ is the thermal average of the neutrino 
annihilation cross section times the relative speed. The three terms on the 
right-hand side of Eq. (\ref{rate}) represent the change in $n$ due
to the expansion of the Universe, the annihilation and creation of the 
massive neutrinos respectively. At the temperatures under consideration, 
the Universe is radiation dominated so that $R$, $T$ and $t$ are related by
\begin{equation}
{{\dot R}\over R} = -{{\dot T}\over T} 
=\left({8\pi \rho \over 3 M_p^2}\right)^{1\over 2},
\label{rad}
\end{equation}
where $M_p$ is the Planck mass. The energy density $\rho$ can be expressed 
in terms of the effective number of degrees of freedom $N_f$ so that
\begin{equation}
\rho = N_f \pi^2 T^4/15,
\label{rho}
\end{equation}
where we have chosen units of temperature such that the Boltzmann constant
is equal to one.
In our case $N_f = 1 + {7 \over 8}(2 + 2)$ (at temperatures less
than $m_{\nu_{\tau}}$) corresponding to the 
photon, two light neutrinos and $e^\pm$ degrees of freedom. By using
Eqs. (\ref{rad}) and (\ref{rho}), the rate equation of Eq. (\ref{rate}) 
can be rewritten in the more convenient form, 
\begin{equation}
{df \over dx} = C \left(f^2 - f_0^2 \right), 
\label{f}
\end{equation}
where
\begin{equation}
C = \left({45 \over 8\pi^3 N_f}\right)^{1/2} m_\nu M_p 
\langle \sigma |v| \rangle, 
\end{equation} 
$x \equiv T/m_{\nu}$ and $f \equiv n/T^3$. To get an approximate analytical 
solution to Eq. (\ref{f}) we use the approximations made in 
Ref. \cite{lswein}; i.e,
\begin{eqnarray}
{df_0 \over dx} & \simeq & C f_0^2, \qquad 
x \simeq x_f \equiv {T_f\over m_\nu}, \nonumber\\ 
{df \over dx} & \simeq & C f^2, \qquad x < x_f,
\label{f2}
\end{eqnarray}
where 
\begin{equation}
f_0 \simeq {g\over (2\pi)^{3\over 2}} x^{-{3\over 2}} e^{-{1\over x}}
\end{equation}
is the equilibrium distribution for nonrelativistic particles
(with $g = 4$ for Dirac fermions) and $T_f$ is the freeze-out temperature.
Solving Eqs. (\ref{f2}) results in
\begin{equation}
x_f^{-{1\over 2}} e^{1\over x_f} \simeq {g\over (2\pi)^{3\over 2}}C
\label{xf}
\end{equation}
and 
\begin{equation}
f(x) \simeq \left[ C\left(x_f^2 + x_f - x \right) \right]^{-1}
\label{fx}
\end{equation}
Therefore the present day mass density of the relic massive tau neutrinos
(including the anti-neutrinos) will be given by
\begin{eqnarray}
\rho_{\nu} & \simeq & m_{\nu} f(0) T^3, \nonumber \\
& = & m_\nu T^3 \left[ C\left(x_f^2 + x_f \right) \right]^{-1},
\end{eqnarray}
where $T$ is evaluated at the present temperature of the relic
photons, $T_{\gamma} = 2.75$ K.  

To get actual numerical estimates, we need to calculate the reaction rate 
$\langle \sigma |v|\rangle$. We find for the s-channel annihilation process,
\begin{equation}
\langle \sigma |v|\rangle
= {2 \pi \delta^2 \alpha^2 \over s^2} 
\beta_e \left\{\left(1 + {1\over 3}\beta^2_e \beta^2_{\nu}\right)s
+ 4\left(m_e^2 + m_{\nu}^2\right)\right\},
\label{cs}
\end{equation}
where $\beta_i = \sqrt{1 - 4m_i^2/s}\ $ ( for $i = e, \nu$ ) and $\alpha$ 
is the electromagnetic fine structure constant. In the non-relativistic 
limit Eq. (\ref{cs}) reduces to:
\begin{equation}
\langle \sigma |v|\rangle \simeq {\delta^2 \alpha^2 \pi \over m_{\nu}^2}
\end{equation}
which is the expression that we will use to evaluate Eqs. (\ref{xf})
and (\ref{fx}). In doing so we obtain
\begin{equation}
{\rho_\nu \over \rho_c} \simeq 
\left({7 \times 10^{-7} \over \delta}\right)^2 
\left({m_\nu \over \hbox{MeV}}\right)^2
{1 \over x_f (x_f + 1)}
\end{equation}
and 
\begin{equation}
{2 \over x_f} + \log{1\over x_f} \simeq 
2\log\left\{\left({\delta \over 3.4 \times 10^{-9} }\right)^2 
\left({m_\nu \over \hbox{MeV}}\right) \right\},
\end{equation}
where $\rho_c = 1.88 \times 10^{-29}h^2$ gcm$^{-3}$ is the
critical density. For simplicity we will take the $h$ parameter
to be one. The requirement that $\rho_\nu / \rho_c \lsim 1 $ gives
a bound on the neutrino's charge and mass. The numerical solutions to 
the above equations are presented in Fig. 1. It is found that for 
$10^{-6} \lsim \delta \lsim 10^{-4}$, $x_f \gsim 0.05$ and so the
energy density bound can approximately be written as
\begin{equation}
{\rho_\nu \over \rho_c} \lsim 
\left({3 \times 10^{-6} \over \delta}\right)^2 
\left({m_\nu \over \hbox{MeV}}\right)^2 \leq 1,
\end{equation}
where $1 < m_\nu \leq 35$ MeV. 

Therefore, if the tau neutrino has a mass greater than 1 MeV, then it can 
annihilate sufficiently in the early Universe provided it has a small 
electric charge. Thus, by endowing the tau neutrino with a small electric 
charge, the first major constraint for consistency with standard cosmology 
can be satisfied.  Also note that the tau neutrino can close the Universe 
and is thus a potential dark matter candidate.

The second constraint from consistency with the standard cosmology model
is that the energy density in the early Universe cannot be too large,
otherwise too much helium will be produced. In our scenario, all of
the charged tau neutrinos will continue to annihilate until the temperature 
drops below about 0.01 to 2.5 MeV (depending on the mass of the tau neutrino)
so that during nucleosynthesis their numbers are heavily depleted relative 
to the number of relativistic species. Thus, our scenario has the novel 
feature that there can be effectively only 2 ( plus a fraction ) neutrinos 
( depending on its mass and charge ) present during nucleosynthesis.

The bound on the effective number of neutrinos present during nucleosynthesis
from standard cosmology depend on some assumptions. For example, the 
authors of Ref. \cite{walk} and Ref. \cite{sar} derive the bounds 
$N_{\nu} \le 3.4$ and $N_{\nu} < 5$ respectively, while some other 
analyses suggest that $N_{\nu}$ could be less than 3 \cite{ri}. One of the 
points of this paper is that a cosmology bound of less than 3 neutrinos 
during nucleosynthesis is not in conflict with particle physics and may 
indicate the possibility of a charged tau neutrino. Also, note that there 
are many models with massive neutrinos (or with light exotic particles) 
which have effectively more than 3 neutrino species present during 
nucleosynthesis. So many of these models have some conflict with the 
standard cosmology model. However, if the tau neutrino is charged,
as discussed above, many of these models can become consistent with
standard cosmology.

\vskip 2cm
\centerline{\bf Figure Caption}
\vskip 1cm
\noindent
Fig. 1. The region of ($m_\nu$, $\delta$) parameter space
allowed by accelerator and cosmology bounds.

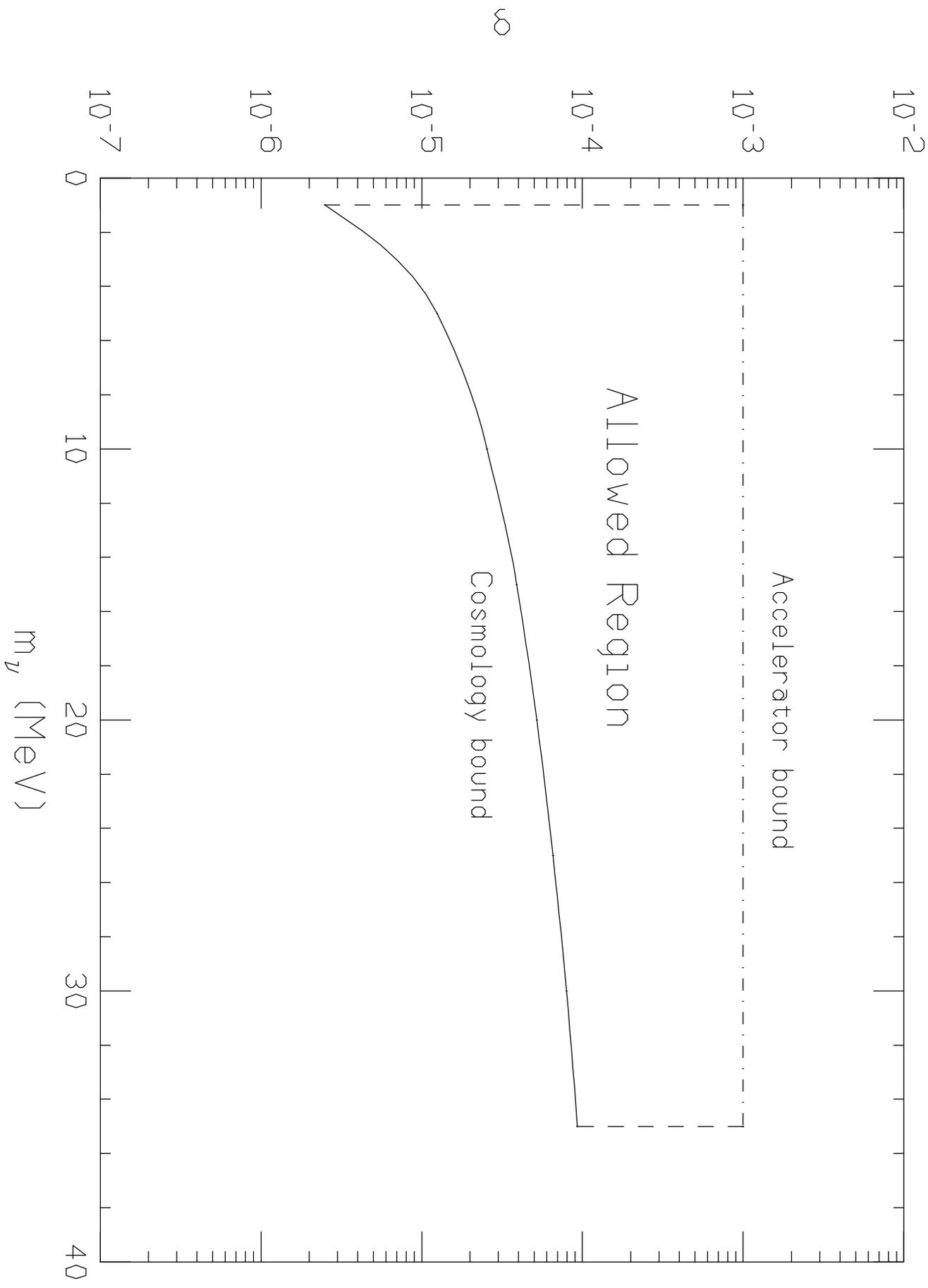
\end{document}